\documentclass[proceedings]{JHEP3} 
\PrHEP{PrHEP unesp2002}

\usepackage{epsfig,multicol}                    
\newbox\mybox
\newcommand\fverb{\setbox\mybox=\hbox\bgroup\verb}
\newcommand\fverbdo{\egroup\medskip\noindent\fbox{\unhbox\mybox}\ }
\newcommand\fverbit{\egroup\item[\fbox{\unhbox\mybox}]}

\title{Conductance from Non-perturbative Methods II}

\author{\speaker{Olalla A. Castro-Alvaredo} and Andreas Fring \\
        Institut f\"ur Theoretische Physik, Freie Universit\"at Berlin,\\
 Arnimallee 14, D-14195 Berlin, Germany\\
        E-mail: \email{Olalla@physik.fu-berlin.de}, 
        \email{Fring@physik.fu-berlin.de}}

\conference{Workshop on Integrable Theories, Solitons and Duality}

\abstract{This talk provides a natural continuation of the talk presented 
by Andreas Fring in this conference. 
Part I was focused on explaining how the DC conductance 
for a free Fermion theory in the presence of different kinds of defects
can be computed by evaluating the Kubo formula. In this talk I will 
focus on an alternative method for the computation of the
same quantity, that is the evaluation of Landauer formula. Once again,
the integrability of the theories under consideration will be exploited, since
a thermodynamic Bethe ansatz analysis provides all the input needed
 in that case, apart from the corresponding reflection and 
transmition amplitudes of the
defect. The basic conclusion of our analysis will be the perfect agreement
between the two different theoretical descriptions mentioned.}

\begin{document} 

The results I will talk about are contained on a series of papers \cite
{CF9,CFF,CF6,CF7,CFG,CF8}, with emphasis on the first two, resulting mainly
from a collaboration with Andreas Fring, who
presented the first part of the work.

\section{Thermodynamic Bethe ansatz for impurity systems}

\noindent As mentioned in the previous talk, for the evaluation of the
Landauer formula \cite{Land} one needs to know the density distribution
functions involved. I will now present a general method which allows to
compute such quantities non-perturbatively, i.e. the thermodynamic Bethe
ansatz (TBA) approach, which we generalized in \cite{CF9} to incorporate the
non-trivial effects arising due to the presence of impurities. \noindent
Besides the aim we have in mind, in general the TBA is a powerful tool for
the computation of thermodynamic quantities in 1+1 dimensional integrable
systems. Originally formulated by Yang and Yang \cite{Yang} in the context
of the non-relativistic Bose gas, it was thereafter generalized by
Zamolodchikov \cite{TBAZam} to relativistic quantum field theories which
interact by means of factorizable scattering matrices. A TBA-analysis serves
to check the consistency of a certain S-matrix proposal, since it allows for
extracting some distinct structural quantities such as the Virasoro central
charge of the underlying conformal field theory. The original bulk
formulation has been accommodated to a situation which includes a purely
transmitting defect in \cite{Marcio}, whereas for purely reflecting
impurities (that is, boundaries) the TBA equations were newly derived in 
\cite{LMSS}. 
\FIGURE{\epsfig{file=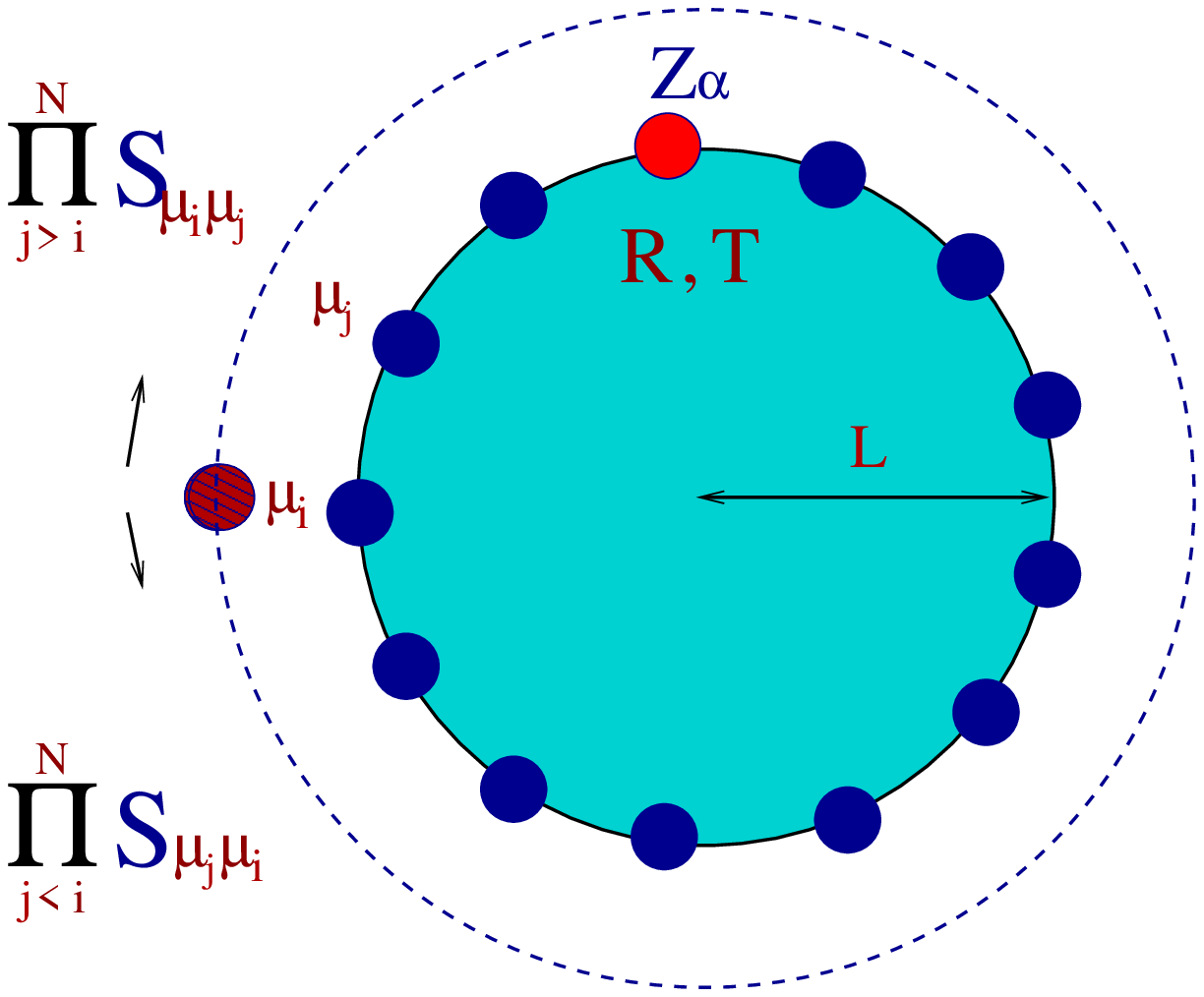,width=8.7cm,height=7.1cm}
   \caption{The Bethe wave function for a defect system.}} \noindent In this
section we want to propose a new formulation which, for the first time, will
hold for a situation when non-vanishing reflection and transmission occur
simultaneously. Let us consider first of all the standard starting point in
the formulation of the Bethe ansatz equations: We consider a 1+1 dimensional
system with compactified space dimension $L$ and $N$ particles distributed
as shown in figure 1. As standard in this context, we represent those
particles by means of the Zamolod- chikov-Faddeev (ZF) operators \cite{FZ} $%
Z_{i}(\theta )$. In addition, in order to incorporate the presence of
boundaries \cite{Chered,Skly,FK} or defects \cite{DMS,CFG} in the system,
the ZF-algebra has to be extended with new generators $Z_{\alpha }$. We
indicate particle types by Latin and degrees of freedom of the impurity by
Greek letters. The ``braiding'' (exchange) relations of annihilation
operators $Z_{i}(\theta )$ of a particle of type $i$ moving with rapidity $%
\theta $ and defect operators $Z_{\alpha }$ in the state $\alpha $ can be
written as 
\begin{eqnarray}
Z_{i}(\theta _{1})Z_{j}(\theta _{2}) &=&S_{ij}^{kl}(\theta _{1}-\theta
_{2})Z_{k}(\theta _{2})Z_{l}(\theta _{1}),  \label{Z0} \\
Z_{i}(\theta _{1})Z_{j}^{\dagger }(\theta _{2}) &=&S_{ij}^{kl}(\theta
_{1}-\theta _{2})Z_{k}^{\dagger }(\theta _{2})Z_{l}(\theta _{1})+2\pi \delta
(\theta _{1}-\theta _{2})\delta _{ij},  \label{Z1/2} \\
Z_{i}(\theta )Z_{\alpha } &=&R_{i\alpha }^{j\beta }(\theta )Z_{j}(-\theta
)Z_{\beta }+T_{i\alpha }^{j\beta }(\theta )Z_{\beta }Z_{j}(\theta )\,,
\label{Z1} \\
Z_{\alpha }Z_{i}(\theta ) &=&\tilde{R}_{i\alpha }^{j\beta }(-\theta
)Z_{\beta }Z_{j}(-\theta )+\tilde{T}_{i\alpha }^{j\beta }(-\theta
)Z_{j}(\theta )Z_{\beta }.  \label{Z2}
\end{eqnarray}
The bulk scattering matrix is indicated by $S$, and the left/right
reflection and transmission amplitudes through the defect are denoted by $R/%
\tilde{R}$ and $T/\tilde{T}$, respectively as seen in part I. We employed
Einstein's sum convention, that is we assume sums over doubly occurring
indices. We suppress the explicit mentioning of the dependence of $Z_{\alpha
}$ on the position in space and assume for the time being that it is
included in $\alpha $. For the treatment of a single defect this is not
relevant, but it will become once more important when we consider multiple
defects. The same relations hold when we replace the annihilation operators
by the creation operators $Z_{i}^{\dagger }(\theta )$ with $R/\tilde{R}$, $T/%
\tilde{T}$ and $S$ replaced by their complex conjugates. The algebra (\ref
{Z1})-(\ref{Z2}) constitutes the starting point for the derivation of the
relations (2.1; part I) and (2.2; part I) which result just from applying it
twice. As usual, we obtain the Bethe ansatz equations by dragging a particle 
$i$ along the world line. We introduce for convenience the following
shorthand notation for the product of various particle operators $%
Z_{i}(\theta )$ and a defect operator $Z_{\alpha }$ 
\begin{equation}
Z_{k,\alpha }^{\mu _{1}\ldots \mu _{N}}:=Z_{\mu _{1}}(\theta _{\mu
_{1}})\ldots Z_{\mu _{k}}(\theta _{\mu _{k}})Z_{\alpha }Z_{\mu
_{k+1}}(\theta _{\mu _{k}+1})\ldots Z_{\mu _{N}}(\theta _{\mu _{N}}).
\label{Zalfa}
\end{equation}
Then we compute the braiding of a particle operator of type $i$ and the
previous product $Z_{k,\alpha }^{\mu _{1}\ldots \mu _{N}}$ by using the
algebra (\ref{Z0})-(\ref{Z2}) and assuming the S-matrix of the bulk theory
to be diagonal 
\begin{eqnarray}
Z_{i}(\theta _{i})Z_{k,\alpha }^{\mu _{1}\ldots \mu _{N}} &=&Z_{k,\alpha
}^{\mu _{1}\ldots \mu _{N}}Z_{i}(\theta _{i})\tilde{F}_{i\alpha
}-Z_{k,\alpha }^{\mu _{1}\ldots \mu _{N}}Z_{i}(-\theta _{i})\tilde{G}%
_{i\alpha }\,,  \label{p1} \\
Z_{k,\alpha }^{\mu _{1}\ldots \mu _{N}}Z_{i}(\theta _{i}) &=&Z_{i}(\theta
_{i})Z_{k,\alpha }^{\mu _{1}\ldots \mu _{N}}F_{i\alpha }-Z_{i}(-\theta
_{i})Z_{k,\alpha }^{\mu _{1}\ldots \mu _{N}}G_{i\alpha }\,.  \label{p2}
\end{eqnarray}
We abbreviated here 
\begin{eqnarray}
\tilde{F}_{i}^{\alpha }\!\!\! &=&\!\!\!\frac{1}{\tilde{T}_{i}^{\alpha
}(-\theta _{i})}\prod_{l=1}^{N}S_{i\mu _{l}}(\theta _{i\mu _{l}})\,,%
\mathnormal{\quad }\tilde{G}_{i}^{\alpha }=\frac{\tilde{R}_{i}^{\alpha
}(-\theta _{i})}{\tilde{T}_{i}^{\alpha }(-\theta _{i})}\prod_{l=1}^{k}S_{i%
\mu _{l}}(\theta _{i\mu _{l}})\prod_{l=k+1}^{N}S_{i\mu _{l}}(-\hat{\theta}%
_{i\mu _{l}})\,,\,\,\,\,\, \\
F_{i}^{\alpha }\!\!\! &=&\!\!\!\frac{1}{T_{i}^{\alpha }(\theta _{i})}%
\prod_{l=1}^{N}S_{\mu _{l}i}(\theta _{\mu _{l}i})\,,\mathnormal{\quad
\,\,\,\,\,}G_{i}^{\alpha }=\frac{R_{i}^{\alpha }(\theta _{i})}{T_{i}^{\alpha
}(\theta _{i})}\prod_{l=1}^{k}S_{\mu _{l}i}(\hat{\theta}_{\mu
_{l}i})\prod_{l=k+1}^{N}S_{\mu _{l}i}(\theta _{\mu _{l}i})\,\,.
\end{eqnarray}
Being on a circle of length $L$, we can make the usual assumption on the
Bethe wavefunction (see e.g. \cite{TBAZam}) which is captured in the
requirement 
\begin{equation}
Z_{i}(\theta )Z_{k,\alpha }^{\mu _{1}\ldots \mu _{N}}=Z_{k,\alpha }^{\mu
_{1}\ldots \mu _{N}}Z_{i}(\theta )\,\exp (-iLm_{i}\sinh \theta )\,.
\end{equation}
Using this monodromy property together with the braiding relations (\ref{p1}%
), (\ref{p2}) and the unitarity relations for $R$ and $T$ (see section 2 of
part I), we obtain the following Bethe ansatz equations 
\begin{equation}
\prod_{l=1}^{N}\frac{S_{li}(\hat{\theta}_{li})}{S_{li}(\theta _{li})}\left(
\prod_{l=1}^{N}S_{li}(\theta _{li})-\frac{e^{iLm_{i}\sinh \theta _{i}}}{%
\tilde{T}_{i}^{\alpha }(-\theta _{i})}\right) =\frac{T_{i}^{\alpha }(-\theta
_{i})}{\tilde{T}_{i}^{\alpha }(-\theta _{i})}\left( \frac{e^{-iLm_{i}\sinh
\theta _{i}}}{T_{i}^{\alpha }(\theta _{i})}-\prod_{l=1}^{N}S_{il}(\theta
_{il})\right) \,.  \label{BA}
\end{equation}
We restrict it here to the diagonal case, i.e. $S_{ij}^{kl}(\theta
)=S_{ij}(\theta )\delta _{li}\delta _{kj},$ $R_{i\alpha }^{j\beta }(\theta
)=R_{i}^{\alpha }(\theta )\delta _{\alpha \beta }\delta _{ij}$, $T_{i\alpha
}^{j\beta }(\theta )=T_{i}^{\alpha }(\theta )\delta _{\alpha \beta }\delta
_{ij}$ and similarly for the tilde amplitudes. We can therefore use the
result mentioned in part I, namely that for $R$ and $T$ to be simultaneously
non-vanishing the only possible bulk scattering matrices are $S=\pm 1$, such
that the relation (\ref{BA}) may be re-written as 
\begin{equation}
1=e^{iLm_{i}\sinh \theta }D_{i\alpha }^{\pm }(\theta
)\prod\nolimits_{l=1}^{N}S_{il}  \label{Ban}
\end{equation}
where 
\begin{equation}
D_{i\alpha }^{\pm }(\theta )=\frac{\tilde{T}_{i}^{\alpha }(\theta
)+T_{i}^{\alpha }(\theta )\prod\nolimits_{l=1}^{N}S_{il}^{2}}{2}\pm \frac{1}{%
2}\left[ \left( \tilde{T}_{i}^{\alpha }(\theta )+T_{i}^{\alpha }(\theta
)\prod\limits_{l=1}^{N}S_{il}^{2}\right) ^{2}-\frac{4T_{i}^{\alpha }(\theta
)\prod\nolimits_{l=1}^{N}S_{il}^{2}}{T_{i}^{\alpha }(-\theta )}\right] ^{%
\frac{1}{2}}.  \label{ds}
\end{equation}
For consistency reasons it is instructive to consider the limit when the
reflection amplitude tends to zero. In that case we can employ the unitarity
relations for the reflection and transmission amplitudes (see section 2 of
part I) and may take the square root in (\ref{ds}), such that we obtain from
(\ref{Ban}) the two equations 
\begin{equation}
R,\tilde{R}\rightarrow 0:\qquad 1=e^{iLm_{i}\sinh \theta }\tilde{T}%
_{i}^{\alpha }(\theta )\prod_{l=1}^{N}S_{il}\,,\mathnormal{\quad }%
1=e^{-iLm_{i}\sinh \theta }T_{i}^{\alpha }(\theta )\prod_{l=1}^{N}S_{li}\,\,.
\label{lba}
\end{equation}
This means we recover the Bethe ansatz equations for a purely transmitting
defect, which were originally proposed by Martins in \cite{Marcio}. The two
signs in (\ref{ds}) capture the breaking of parity invariance in the
limiting case, i.e. the two equations in (\ref{lba}) correspond to taking
the particle either clockwise or anti-clockwise around the world line as
formulated for the parity breaking case for the first time in \cite{CFKM}
and explicitly indicated in figure 1. We do not expect to recover from here
the equations for a purely reflecting boundary which were suggested in \cite
{LMSS}, since the equations (\ref{p1}) and (\ref{p2}) do not make sense in
the limit $T,\tilde{T}\rightarrow 0$. For $\prod%
\nolimits_{l=1}^{N}S_{il}^{2}=1,$ i.e. the free Boson and Fermion, we can
exploit the fact that (\ref{Ban}) with (\ref{ds}) look formally precisely
like the Bethe ansatz equations for a purely transmitting defect. If we want
to exploit this analogy we should of course be concerned about the question
whether $D_{j\alpha }^{\pm }(\theta )$ is a meromorphic function. Assuming
parity invariance, we may take the square root \quad 
\begin{equation}
D_{j\alpha }^{\pm }(\theta )=T_{j}^{\alpha }(\theta )\,\pm R_{j}^{\alpha
}(\theta )\quad \mathnormal{for}\quad R=\tilde{R},\,T=\tilde{T}.  \label{s}
\end{equation}
The matrix $D_{j\alpha }^{\pm }(\theta )$ has now the usual properties,
namely it is unitarity in the sense that $D_{j\alpha }^{\pm }(\theta
)D_{j\alpha }^{\pm }(-\theta )=1$. It follows further from (\ref{s}) and
from the crossing relations for $R$ and $T$ that the hermiticity relation $%
D_{j\alpha }^{\pm }(\theta )=D_{j\alpha }^{\pm }(-\theta )^{\ast }$ and the
crossing relations $D_{\bar{\jmath}\alpha }^{\pm }(\theta )=D_{j\alpha
}^{\mp }(i\pi -\theta )$ and $D_{\bar{\jmath}\alpha }^{\pm }(\theta
)=D_{j\alpha }^{\pm }(i\pi -\theta )$ hold for the free Fermion and Bosons,
respectively.

Let us now carry out the thermodynamic limit in the usual way, namely by
increasing the particle number $N$ and the system size $L$ in such a way
that their mutual ratio $N/L$ remains finite. The amount of defects will be
kept constant in this limit, such that there is no contribution to the
TBA-equations from the defect in that situation. The same behaviour was
pointed out in \cite{Marcio} for the purely transmitting case. Intuitively
the latter result was to be expected, since making both the amount of
particles and the size of the system infinite while keeping the amount of
defects fixed will lead to a situation in which the effect of the presence
of a finite number of defects is negligeable. Hence, this means that
essentially we can employ the usual bulk TBA analysis when the
considerations are carried out in the thermodynamic limit.

Let us therefore recall the main equations of the TBA analysis. For more
details on the derivation see \cite{TBAZam} and in particular for the
introduction of the chemical potential see \cite{TBAKM}. The main input into
the entire analysis is the dynamical interaction, which enters via the
logarithmic derivative of the scattering matrix $\varphi _{ij}(\theta
)=-id\ln S_{ij}(\theta )/d\theta $ and the assumption on the statistical
interaction, which we take to be Fermionic. As usual \cite{TBAZam,TBAKM}, we
take the logarithmic derivative of the Bethe ansatz equation (\ref{Ban}) and
relate the density of states $\rho _{i}(\theta ,r)$ for particles of type $i$
as a function of the inverse temperature $r=1/T$ to the density of occupied
states $\rho _{i}^{r}(\theta ,r)$ 
\begin{equation}
\rho _{i}(\theta ,r)=\frac{m_{i}}{2\pi }\cosh \theta
+\sum\nolimits_{j}[\varphi _{ij}\ast \rho _{i}^{r}](\theta )\,.  \label{rho}
\end{equation}
By $\left( f\ast g\right) (\theta )$ $:=1/(2\pi )\int d\theta ^{\prime
}f(\theta -\theta ^{\prime })g(\theta ^{\prime })$ we denote as usual the
convolution of two functions. The mutual ratio of the densities serves as
the definition of the so-called pseudo-energies $\varepsilon _{i}(\theta ,r)$%
\begin{equation}
\frac{\rho _{i}^{r}(\theta ,r)}{\rho _{i}(\theta ,r)}=\frac{e^{-\varepsilon
_{i}(\theta ,r)}}{1+e^{-\varepsilon _{i}(\theta ,r)}}\,,  \label{dens}
\end{equation}
which have to be positive and real. Notice that, from the introduction in
part I, the quantities $\rho _{i}^{r}(\theta ,r)$ \ at the constrictions of
the wire are the basic input we need for the evaluation of Landauer formula,
apart from the reflection and transmission amplitudes. At thermodynamic
equilibrium one obtains then the TBA-equations, which read in these
variables and in the presence of a chemical potential $\mu _{i}$%
\begin{equation}
rm_{i}\cosh \theta =\varepsilon _{i}(\theta ,r,\mu _{i})+r\mu
_{i}+\sum\nolimits_{j}[\varphi _{ij}\ast \ln (1+e^{-\varepsilon
_{j}})](\theta )\,,  \label{TBA}
\end{equation}
where $r=m/T$, $m_{l}\rightarrow m_{l}/m$, $\mu _{i}\rightarrow \mu _{i}/m$,
with $m$ being the mass of the lightest particle in the model. It is
important to note that $\mu _{i}$ is restricted to be smaller than 1. This
follows immediately from (\ref{TBA}) by recalling that $\varepsilon _{i}\geq
0$ and that for $r$ large $\varepsilon _{i}(\theta ,r,\mu _{i})$ tends to
infinity. As pointed out already in \cite{TBAZam} (here just with the small
modification of a chemical potential), the comparison between (\ref{TBA})
and (\ref{rho}) leads to the useful relation 
\begin{equation}
\rho _{i}(\theta ,r,\mu _{i})=\frac{1}{2\pi }\left( \frac{d\varepsilon
_{i}(\theta ,r,\mu _{i})}{dr}+\mu _{i}\right) \,.  \label{rhoe}
\end{equation}
The main task is therefore first to solve (\ref{TBA}) for the
pseudo-energies from which then all densities can be reconstructed.

\subsection{Thermodynamic quantities per unit length\ }

\noindent Treating the equations (\ref{Ban}) and (\ref{ds}) in the mentioned
analogy with the purely transmitting case we can also construct various
thermodynamic quantities. It should be stressed that these quantities are
computed per unit length. Similarly as the expression found in \cite{Marcio}
for a purely transmitting defect the free energy is 
\begin{equation}
F(r)=-\frac{1}{\pi r}\sum_{l,\alpha }\hat{m}_{l}\int\nolimits_{0}^{\infty
}d\theta \,[\cosh \theta +m^{-1}\varphi _{l\alpha }(\theta )]\,\ln [1+\exp
(-rm\cosh \theta )]\,.  \label{scale}
\end{equation}
It is made up of two parts, one coming from the bulk and one including the
data of the defect in form of $\varphi _{l\alpha }(\theta )=-id\ln
D_{l\alpha }(\theta )/d\theta $. From equation (\ref{scale}) we also see
that when taking the mass scale to be large in comparison to the dominating
scale in the defect, the latter contribution to the scaling function becomes
negligible with regard to the bulk and vice versa. However, in this talk we
shall concentrate on the thermodynamic limit.

\section{Conductance through an impurity}

\noindent The most intuitive way to compute the conductance is via
Landauer-B\"{u}ttinger transport theory \cite{Land}. Let us consider a set
up as depicted in figure 2, that is we place a defect in the middle of a
rigid bulk wire, where the two halves might be at different temperatures. 
The direct current $I$ through such a quantum wire can be computed simply by
determining the difference between the static charge distributions at the
right and left constriction of the wire, i.e. $I=Q_{r}-Q_{l}$. This is based
on the assumption \cite{FLS,LSS}, that $Q(t)\sim (Q_{r}-Q_{l})t\sim $ $(\rho
_{r}-\rho _{l})t$, where the $\rho $s are the corresponding density
distribution functions. Placing an

\begin{center}
\FIGURE{\epsfig{file=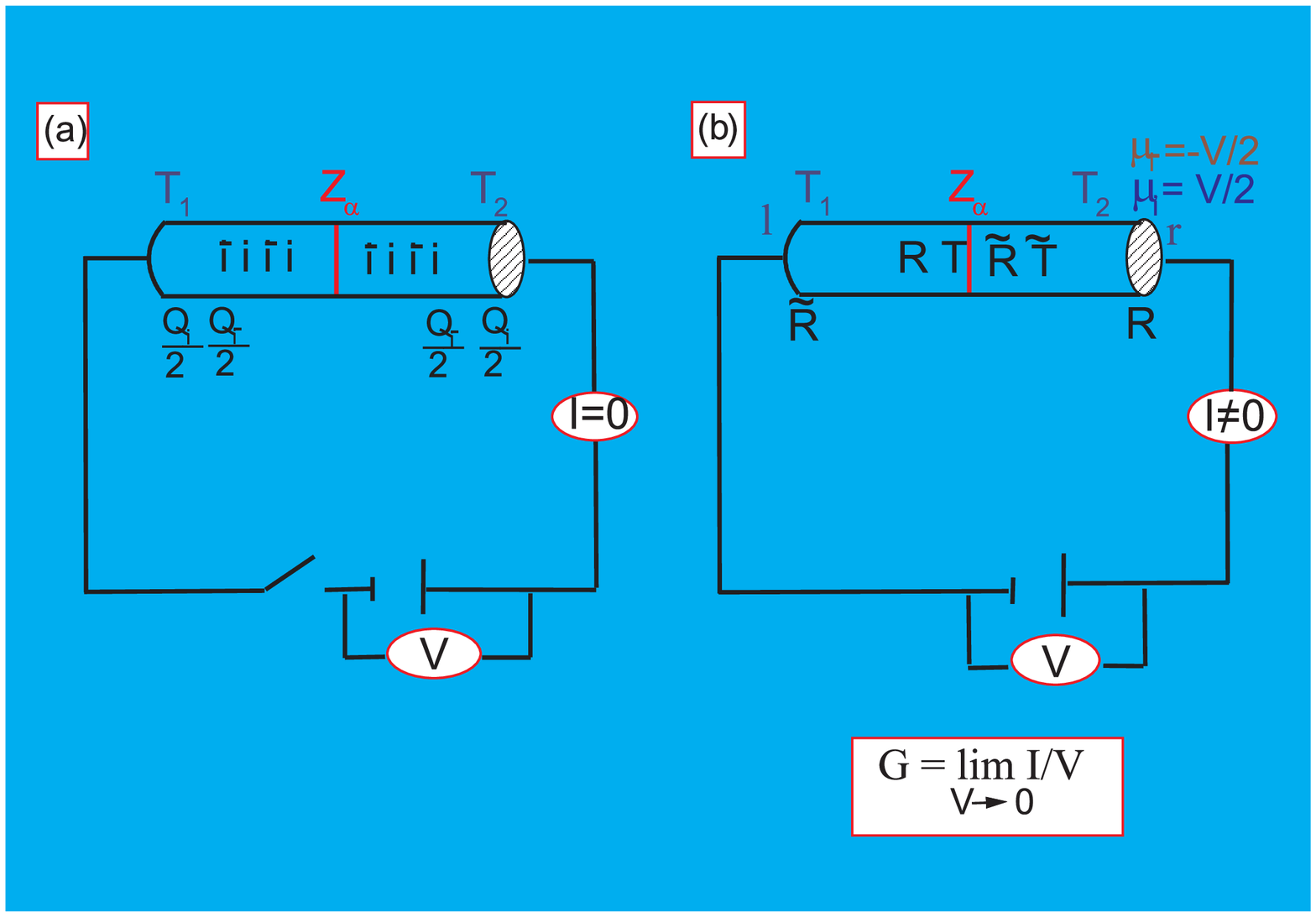,width=9.4cm,height=5.9cm}
        \caption{A conductance measurement. Part (a) represents
the initial condition with no current flowing, i.e., I=0 and part (b), $I\neq 0$. 
The defect is placed in the middle of the wire and the left and
right half are assumed to be at temperatures $T_{1}$ and $T_{2}$,
respectively.}}
\end{center}

\noindent impurity in the middle of the wire, we have to quantify the
overall balance of particles of type $i$ and anti-particles $\bar{\imath}$
carrying opposite charges $q_{i}=-q_{\bar{\imath}}\,$\ at the end of the
wire at different potentials. This information is of course encoded in the
density distribution function $\rho _{i}^{r}\left( \theta ,T,\mu _{i}\right) 
$.$\,$In the described set up half of the particles of one type are already
at the same potential at one of the ends of the wire and the probability for
them to reach the other is determined by the transmission and reflection
amplitudes through the impurity. We assume that there is no effect coming
from the constrictions of the wire, i.e. they are purely transmitting
surfaces with $T=\tilde{T}=1$. One could, however, also consider a situation
in which those constrictions act as boundaries, namely purely reflecting
surfaces. The situation could be described with the same transport theory
picture, see e.g. \cite{FLS,FLS2,casa}, but then the conductance can only be
non-vanishing if the reflection amplitudes in the constrictions are
non-diagonal in the particle degrees of freedom, such as for instance for
sine-Gordon \cite{GZ}, that is in general affine Toda field theories with
purely imaginary coupling constant or, in the massless limit, folded purely
reflecting (transmitting) diagonal bulk theories.

According to the Landauer transport theory the direct current (DC) along the
wire is given by 
\begin{eqnarray}
I^{\mathbf{\vec{\alpha}}} &=&\sum_{i}I_{i}^{\mathbf{\vec{\alpha}}}(r,\mu
_{i}^{l},\mu _{i}^{r})=\sum_{i}\frac{q_{i}}{2}\int\limits_{-\infty }^{\infty
}d\theta \left[ \rho _{i}^{r}(\theta ,r,\mu _{i}^{r})|T_{i}^{\mathbf{\vec{%
\alpha}}}\left( \theta \right) |^{2}-\rho _{i}^{r}(\theta ,r,\mu _{i}^{l})|%
\tilde{T}_{i}^{\mathbf{\vec{\alpha}}}\left( \theta \right) |^{2}\right]
,\quad \,\,\,\,\,\,\,  \label{I11} \\
&=&I_{B}-\sum_{i}\frac{q_{i}}{2}\int\limits_{-\infty }^{\infty }d\theta %
\left[ \rho _{i}^{r}(\theta ,r,\mu _{i}^{r})|R_{i}^{\mathbf{\vec{\alpha}}%
}\left( \theta \right) |^{2}-\rho _{i}^{r}(\theta ,r,\mu _{i}^{l})|\tilde{R}%
_{i}^{\mathbf{\vec{\alpha}}}\left( \theta \right) |^{2}\right] \,\,,
\label{I2}
\end{eqnarray}
where we assume here $T_{1}=T_{2}$. The relation (\ref{I2}) is obtained from
(\ref{I11}) simply by making use of the fact that $|R|^{2}+|T|^{2}=1$ (see
section 2 in part I). Equation (\ref{I2}) has the virtue that it extracts
explicitly the bulk contribution to the current which we refer to as $I_{B}$%
. There are some obvious limits, namely a transparent and an impenetrable
defect 
\begin{equation}
\lim_{|T^{\mathbf{\vec{\alpha}}}|\rightarrow 1}I^{\mathbf{\vec{\alpha}}%
}=I_{B}\qquad \mathnormal{and}\qquad \lim_{|T^{\mathbf{\vec{\alpha}}%
}|\rightarrow 0}I^{\mathbf{\vec{\alpha}}}=0\,,
\end{equation}
respectively. A short comment is needed on the validity of (\ref{I11}).
Apparently it suggests that when the parity between left and right
scattering is broken, there is the possibility of a net current even when an
external source is absent. In this picture we have of course not taken into
account that charged particles moving through the defect will alter the
potential, such that we did in fact not describe a perpetuum mobile. Thus
the limitation of our analysis is that $\mu _{i}^{l}-\mu _{i}^{r}$ has to be
much larger than the change in the potential induced by the moving particles.

Finally we want to compute the conductance from the DC current, which by
definition is obtained from 
\begin{equation}
G^{\mathbf{\vec{\alpha}}}(r)=\sum\nolimits_{i}G_{i}^{\mathbf{\vec{\alpha}}%
}(r)=\sum\nolimits_{i}\lim_{(\mu _{i}^{l}-\mu _{i}^{r})\rightarrow 0}I_{i}^{%
\mathbf{\vec{\alpha}}}(r,\mu _{i}^{l},\mu _{i}^{r})\,/(\mu _{i}^{l}-\mu
_{i}^{r})  \label{G}
\end{equation}
and is of course a property of the material itself and a function of the
temperature. In general the expressions in (\ref{I11}) tend to zero for
vanishing chemical potential difference such that the limit in (\ref{G}) is
non-trivial.

Thus from the knowledge of the transmission matrix and the density
distribution function we can compute the conductance.

\subsection{The high temperature regime}

\noindent Since the physical quantities require a solution of the
TBA-equations, which up to now, due to their non-linear nature, can only be
solved numerically, we have to resort in general to a numerical analysis to
obtain the conductance for some concrete theories. However, there exist
various approximations for different special situations, such as the high
temperature regime. For large rapidities and small $r$, it is known \cite
{TBAZam} (here we only need the small modification of the introduction of a
chemical potential $\mu _{i}$) that the density of states can be
approximated by 
\begin{equation}
\rho _{i}(\theta ,r,\mu _{i})\sim \frac{m_{i}}{4\pi }e^{|\theta |}\sim \frac{%
1}{2\pi r}\epsilon (\theta )\frac{d\varepsilon _{i}(\theta ,r,\mu _{i})}{%
d\theta }\,,  \label{rr}
\end{equation}
where $\epsilon (\theta )=\Theta (\theta )-\Theta (-\theta )$ is the step
function, i.e. $\epsilon (\theta )=1$ for $\theta >0$ and $\epsilon (\theta
)=-1$ for $\theta <0$. In equation (\ref{dens}), we assume that in the large
rapidity regime $\rho _{i}^{r}(\theta ,r,\mu _{i})$ is dominated by (\ref{rr}%
) and in the small rapidity regime by the Fermi distribution function.
Therefore 
\begin{equation}
\rho _{i}^{r}(\theta ,r,\mu _{i})\sim \frac{1}{2\pi r}\epsilon (\theta )%
\frac{d}{d\theta }\ln \left[ 1+\exp (-\varepsilon _{i}(\theta ,r,\mu _{i}))%
\right] \,\,.
\end{equation}
Using this expression in equation (\ref{I11}), we approximate the direct
current in the ultraviolet by 
\begin{equation}
\lim\limits_{r\rightarrow 0}I_{i}^{\mathbf{\vec{\alpha}}}(r,\mu _{i})\sim 
\frac{q_{i}}{4\pi r}\int\limits_{-\infty }^{\infty }d\theta \ln \left[ \frac{%
1+\exp (-\varepsilon _{i}(\theta ,r,\mu _{i}^{l}))}{1+\exp (-\varepsilon
_{i}(\theta ,r,\mu _{i}^{r}))}\right] \,\frac{d\left[ \epsilon (\theta
)\,|T_{i}^{\mathbf{\vec{\alpha}}}(\theta )|^{2}\right] }{d\theta }\,,
\label{ya}
\end{equation}
after a partial integration. For simplicity we also assumed here parity
invariance, that is $|T_{i}^{\mathbf{\alpha }}(\theta )|=|\tilde{T}_{i}^{%
\mathbf{\alpha }}(\theta )|$. The derivation of the analogue to (\ref{ya})
for the situation when parity is broken is of course similar. Taking now the
potentials at the end of the wire to be $\mu _{i}^{r}=-\mu _{i}^{l}=V/2$,
the conductance reads in this approximation 
\begin{equation}
\lim\limits_{r\rightarrow 0}G_{i}^{\mathbf{\vec{\alpha}}}(r)\sim \frac{q_{i}%
}{2\pi r}\int\limits_{-\infty }^{\infty }d\theta \frac{1}{1+\exp
[\varepsilon _{i}(\theta ,r,0)]}\left. \frac{d\varepsilon _{i}(\theta ,r,V/2)%
}{dV}\right| _{V=0}\frac{d\left[ \epsilon (\theta )\,|T_{i}^{\mathbf{\vec{%
\alpha}}}(\theta )|^{2}\right] }{d\theta }\,.
\end{equation}
In order to evaluate these expressions further, we need to know explicitly
the precise form of the transmission matrix, i.e. the concrete form of the
defect. An interesting situation occurs when the defect is transparent or
rapidity independent, that is $|T_{i}^{\mathbf{\vec{\alpha}}}(\theta
)|\rightarrow |T_{i}^{\mathbf{\vec{\alpha}}}|$, in which case we can pursue
the analysis further. Noting that $d\epsilon (\theta )/d\theta =2\delta
(\theta )$, we obtain 
\begin{equation}
\lim\limits_{r\rightarrow 0}G_{i}^{\mathbf{\vec{\alpha}}}(r)\sim \frac{q_{i}%
}{\pi r}\frac{|T_{i}^{\mathbf{\vec{\alpha}}}|^{2}}{1+\exp \varepsilon
_{i}(0,r,0)}\left. \frac{d\varepsilon _{i}(0,r,V/2)}{dV}\right| _{V=0}\,.
\label{g0}
\end{equation}
The derivative $d\varepsilon _{i}(0,r,V/2)/dV$ can be obtained by solving
recursively 
\begin{equation}
\frac{d\varepsilon _{i}(0,r,V/2)}{dV}=-\frac{r}{2}-\sum_{j}N_{ij}\frac{1}{%
1+\exp \varepsilon _{j}(0,r,V/2)]}\frac{d\varepsilon _{j}(0,r,V/2)}{dV}\,,
\end{equation}
which results form a computation similar to a standard one in this context 
\cite{TBAZam} leading to the so-called constant TBA-equations. Here only the
asymptotic phases of the scattering matrix enter via $N_{ij}=\lim_{\theta
\rightarrow \infty }[\ln [S_{ij}(-\theta )/S_{ij}(\theta )]]/2\pi i$. The
values of $\varepsilon _{i}(0,r,0)$ needed in (\ref{g0}) can be obtained for
small $r$ in the usual way from the standard constant TBA-equations.

\subsection{Free Fermion with defects\ \ }

\noindent Let us exemplify the general formulae once more with the free
Fermion. First of all we note that in this case in the TBA-equations (\ref
{TBA}) the kernel $\varphi _{ij}$ is vanishing and the equation is simply
solved by 
\begin{equation}
\varepsilon _{i}(\theta ,r,\mu _{i})=rm_{i}\cosh \theta -r\mu _{i}\,.
\label{ffe}
\end{equation}
Therefore, we have explicit functions for the densities with (\ref{rhoe})
and (\ref{dens}) 
\begin{equation}
\rho _{i}\left( \theta ,r,\mu _{i}\right) =\frac{1}{2\pi }m_{i}\cosh \theta
\qquad \mathnormal{and\qquad }\rho _{i}^{r}\left( \theta ,r,\mu _{i}\right) =%
\frac{m_{i}\cosh \theta /2\pi }{1+\exp (rm_{i}\cosh \theta -r\mu _{i})}\,.
\label{rfe}
\end{equation}
According to (\ref{I11}) the direct current reads 
\begin{eqnarray}
I^{\mathbf{\vec{\alpha}}}(r,V) &=&\frac{q_{i}}{2}\int\limits_{-\infty
}^{\infty }d\theta \left[ \rho _{\bar{\imath}}^{r}\left( \theta
,r,V/2\right) |T_{\bar{\imath}}^{\mathbf{\vec{\alpha}}}\left( \theta \right)
|^{2}-\rho _{i}^{r}\left( \theta ,r,-V/2\right) |T_{i}^{\mathbf{\vec{\alpha}}%
}\left( \theta \right) |^{2}\right.  \nonumber \\
&&\left. \quad \qquad \quad -\rho _{\bar{\imath}}^{r}\left( \theta
,r,-V/2\right) |\tilde{T}_{\bar{\imath}}^{\mathbf{\vec{\alpha}}}\left(
\theta \right) |^{2}+\rho _{i}^{r}\left( \theta ,r,V/2\right) |\tilde{T}%
_{i}^{\mathbf{\vec{\alpha}}}\left( \theta \right) |^{2}\right] \,\,.
\end{eqnarray}
Using atomic units $m_{e}=e=%
h\hskip-.2em\llap{\protect\rule[1.1ex]{.325em}{.1ex}}\hskip.2em%
=m_{i}=q_{i}=1$, we obtain explicitly with (\ref{rfe}) 
\begin{equation}
I^{\mathbf{\vec{\alpha}}}(r,V)=\frac{1}{\pi }\int\limits_{0}^{\infty
}d\theta \frac{\cosh \theta \sinh (rV/2)\,|T^{\mathbf{\vec{\alpha}}}\left(
\theta \right) |^{2}}{\cosh (r\cosh \theta )+\cosh (rV/2)},  \label{44}
\end{equation}
for $|T_{\bar{\imath}}^{\mathbf{\alpha }}\left( \theta \right) |=|T_{i}^{%
\mathbf{\vec{\alpha}}}\left( \theta \right) |=|\tilde{T}_{\bar{\imath}}^{%
\mathbf{\vec{\alpha}}}\left( \theta \right) |=|\tilde{T}_{i}^{\mathbf{\alpha 
}}\left( \theta \right) |=|T^{\mathbf{\vec{\alpha}}}\left( \theta \right)
|\, $. Then by (\ref{G}) the conductance results to 
\begin{equation}
G^{\mathbf{\vec{\alpha}}}(r)=rm\frac{e^{2}}{h}\int\limits_{0}^{\infty
}d\theta \frac{\cosh \theta \,\left| T^{\mathbf{\vec{\alpha}}}\left( \theta
\right) \right| ^{2}}{1+\cosh (rm\cosh \theta )}\,  \label{gg}
\end{equation}
in this case. We have re-introduced dimensional quantities instead of atomic
units to be able to match with some standard results from the literature.
The most characteristic features can actually be captured when we carry out
the massless limit as indicated in section 2.3.2, which can be done even
analytically. Substituting $t=e^{\theta }$, we obtain 
\begin{equation}
\lim_{m\rightarrow 0}G^{\mathbf{\vec{\alpha}}}(r)\sim \frac{e^{2}}{h}%
\int\limits_{0}^{\infty }dt\frac{\,|T_{L/R}^{\mathbf{\vec{\alpha}}%
}(t\,y/r)|^{2}}{1+\cosh (t)}=\frac{e^{2}}{h}\left\{ 
\begin{array}{l}
\overline{|T_{L/R}^{\mathbf{\vec{\alpha}}}(t\,y/r)|^{2}}\qquad \quad \,\,\,\,%
\mathrm{\,for}\quad y\gg r \\ 
|T_{L/R}^{\mathbf{\vec{\alpha}}}(y/r=0)|^{2}\quad \quad \,\,\mathrm{for}%
\quad y\ll r
\end{array}
\right. \,.  \label{massl}
\end{equation}
We have identified here two distinct regions. When $y\ll r$ we can replace
the left/right transmission amplitudes by their values at $y/r=0$. When $%
y\gg r$ the transmission amplitudes enter the expression as a strongly
oscillatory function in which $y/r$ plays the role of the frequency. It is
then a good approximation to replace this function by its mean value as
indicated by the overbar. It is straightforward to extend the expression (%
\ref{massl}) to the case when the assumption on $T^{\mathbf{\alpha }}$ in (%
\ref{44}) is relaxed and to the case with different values of $y$. To
proceed further we need to specify the defect.

\subsubsection{Transparent defects, $|T^{\mathbf{\vec{\protect\alpha}}}|=1$}

\noindent Let us first consider the easiest example, which supports the
general working of the method. When the defect is transparent, i.e., $|T^{%
\mathbf{\vec{\alpha}}}|=1$, we can compute the expression for the
conductance (\ref{gg}) directly in the large temperature limit and obtain
the well known behaviour \cite{KF} 
\begin{equation}
\lim_{r\rightarrow 0,|T^{\mathbf{\vec{\alpha}}}|\rightarrow 1}G^{\mathbf{%
\vec{\alpha}}}(r)\sim \frac{e^{2}}{h}(1-\frac{rm}{2})\,.  \label{hh}
\end{equation}
Alternatively, we obtain the expression (\ref{hh}) also from equation (\ref
{g0}) and (\ref{ffe}). In the massless limit of (\ref{massl}) we obtain $%
e^{2}/h$ which coincides with the result in \cite{FLS}. However, we should
stress that we consider here purely massive cases and the massless limit
only serves as a benchmark. Note that a transparent defect in this context
does not necessarily mean the absence of the defect, since the transmission
amplitude could be a non-trivial phase.

\subsubsection{The energy operator defect $\mathcal{D}^{\protect\alpha }(%
\bar{\protect\psi},\protect\psi )=g\bar{\protect\psi}\protect\psi $}

\noindent For this defect the computation of the conductance according to (%
\ref{gg}) is more involved. The results of our numerical analysis of the
expression (\ref{gg}) are depicted in figure 3.

\FIGURE{\epsfig{file=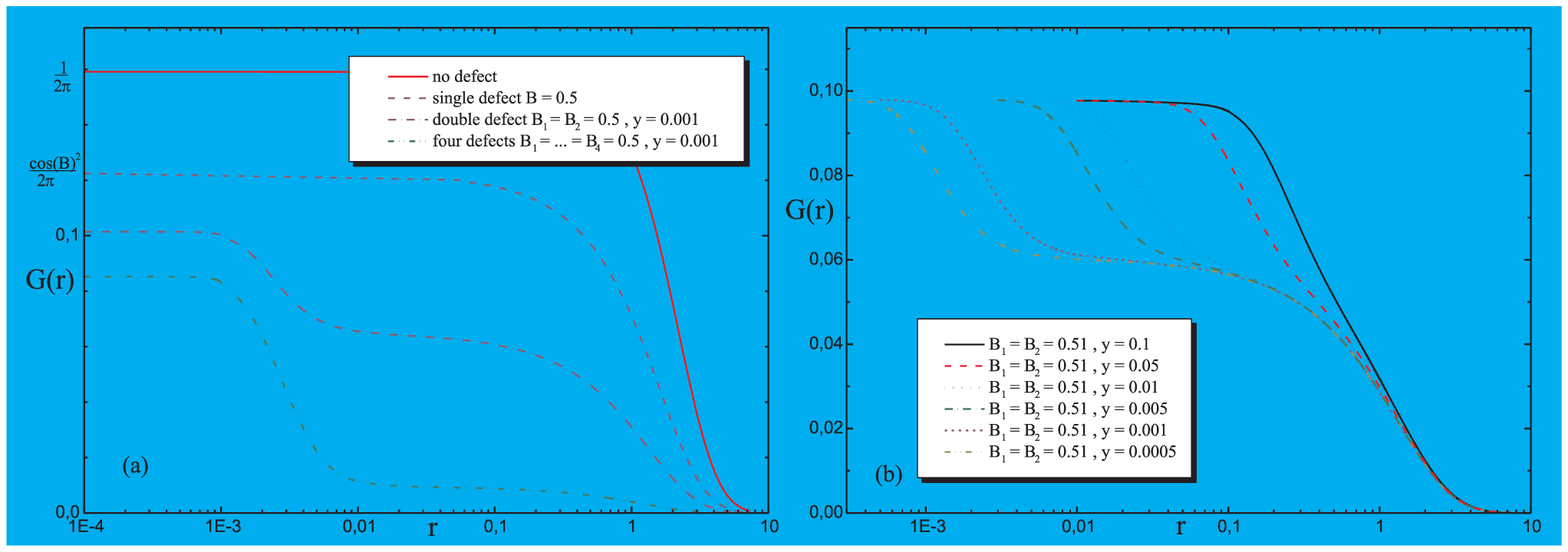,width=15.cm,height=5.7cm}
\caption{Conductance $G(r)$ for the complex
free Fermion with the energy operator defects as a function of the inverse
temperature $r,$  for fixed effective coupling constant $B$ 
 and (a) for varying amounts of defects $\ell =0,1,2,4$. (b) for $\ell =2$
for varying distances $y$.}}

\noindent We observe several distinct features. First of all it is naturally
to be expected that when we increase the number of defects the resistance
will grow. This is confirmed, as for fixed temperature and increasing number
of defects, the conductance decreases. Second we see several well extended
plateaux. They can be reproduced with the analytical expressions obtained in
the massless limit (\ref{massl}). To be able to compare with (\ref{gg}) we
re-introduce atomic units for convenience, i.e. $e^{2}/h\rightarrow 1/2\pi $%
. For a single defect there is only one plateau and from (\ref{massl}) and
the explicit expression for $T^{\alpha }(\theta )$ given in part I 
\begin{equation}
G^{\alpha }(r)\sim \frac{\cos ^{2}B}{2\pi }\,.  \label{g1}
\end{equation}
For $B=0.5$ the value $0.1226$ is well reproduced in figure 3(a). The lower
lying plateaux correspond to the region when $y\ll r$. In that case we
obtain from (\ref{massl}) together with the expressions for the reflection
and transmission amplitudes of the double and four defect systems derived in
part I 
\begin{eqnarray}
G^{\alpha _{1}\alpha _{2}}(r) &\sim &\frac{1}{2\pi }\left( \frac{\cos ^{2}B}{%
1+\sin ^{2}B}\right) ^{2}\qquad \quad \,\quad \quad \qquad \quad \qquad \,{%
\,\,\mathrm{for}\quad }y\ll r,  \label{o1} \\
G^{\alpha _{1}\alpha _{2}\alpha _{3}\alpha _{4}}(r) &\sim &\frac{1}{2\pi }%
\left( \frac{\cos ^{4}B}{\cos ^{4}B-2(1+\sin ^{2}B)^{2}}\right) ^{2}\,\quad
\quad \qquad \,\mathrm{\,for}\quad y\ll r.  \label{o2}
\end{eqnarray}
For $B=0.5$ the values $0.0624$ and $0.0095$ are well reproduced in figure
3(a) for $\ell =2$ and $\ell =4$, respectively. The plateaux extending to
the ultraviolet regime result from (\ref{massl}) and by taking mean values
of the expressions for the reflection and transmission amplitudes of the
double and four defect systems given in part I 
\begin{eqnarray}
G^{\alpha _{1}\alpha _{2}}(r) &\sim &\frac{2}{\pi }\frac{1+\sin ^{4}B}{(\cos
^{2}(2B)-3)^{2}}\,,\qquad \quad \,\quad \qquad \quad \,\,\,\,\,\,\,\,\quad 
\mathrm{for}\qquad y\gg r\,,  \label{o3} \\
G^{\alpha _{1}\alpha _{2}\alpha _{3}\alpha _{4}}(r) &\sim &\frac{1}{4\pi }+%
\frac{\cos ^{8}B}{4\pi \lbrack \cos ^{4}B-2(1+\sin ^{2}B)^{2}]^{2}}%
,\,\,\quad \,\,\quad \mathrm{for}\qquad y\gg r.  \label{o4}
\end{eqnarray}
Also in this case the values for $B=0.5$, i.e., $0.110784$ and $0.084311$
for $\ell =2$ and $\ell =4$, respectively, match very well with the
numerical analysis. Finally we have to explain the reason for the increase
from one to the next plateaux and why the curves are shifted precisely in
the way as indicated in figure 3(b) when we change the distance between the
defects. This phenomenon is attributed to resonances, namely the existence
of very sharp picks in the probability of transmission for two or more
defects (see figure 1 in part I).

If we now compare the expressions (\ref{g1})-(\ref{o4}) with equations
(3.34; part I)-(3.42; part I), we find complete agreement. This observation
constitutes the central result of this work. \emph{We showed for a concrete
integrable theory that the conductance computed by means of the newly
formulated Kubo \cite{CF9,kubo} formula incorporating the presence of
defects, and by means of Landauer \cite{Land} formula are in perfect
agreement.} More concrete examples of this agreement are provided in \cite
{CF9}.

\subsubsection{The $SU(3)_{2}$ homogeneous sine-Gordon model, unstable
particles}

\noindent The $SU(3)_{2}$ homogeneous sine-Gordon (HSG) model is the
simplest of its kind and contains only two self-conjugate solitons, which we
denote by ``+'', ``$-$'', and one unstable particle, which we call $\tilde{c}
$. The corresponding scattering matrix was found \cite{HSGS} to be 
\begin{equation}
S_{\pm \pm }=-1,\qquad S_{\pm \mp }(\theta )=\pm \tanh \frac{1}{2}\left(
\theta \pm \sigma -\frac{i\pi }{2}\right) ,  \label{ss}
\end{equation}
which means the resonance pole associated to the formation $\tilde{c}$ is
situated at $\theta _{R}=\mp \sigma -i\pi /2$, $\sigma $ being a free
parameter. Stable bound states may not be formed. Since only for $S=\pm 1$
simultaneous reflection and transmission can occur \cite{CFG}, the $%
SU(3)_{2} $-HSG model only admits the presence of purely reflecting or
transmitting defects. For the purely reflecting case, the expression (\ref
{I11}) vanishes so that the only non-trivial situation we can consider is a
transparent defect, i.e. $|T|=1$. The results for the conductance after
solving numerically the TBA equations (\ref{rho}) and (\ref{TBA}) are
depicted in figure 4. When solving (\ref{rho}) and (\ref{TBA}) we have taken 
$\mu _{R}=-\mu _{L}=0.25.$ However, according to the definition (\ref{G}) we
should really consider the limit $\left( \mu _{R}-\mu _{L}\right)
\rightarrow 0$. The reason why we instead take $\mu _{R}-\mu _{L}=0.5$ is
that for this model we can of course not solve the TBA-equations
analytically, as for the free Fermion. On the contrary, the numerics become
fairly involved and they do not allow for considering the extreme limit $%
\left( \mu _{R}-\mu _{L}\right) \rightarrow 0$. However, we convinced
ourselves that the results depicted in figure 4 reproduce indeed the correct
behaviour of the conductance, since computing $G(r)$ in the deep ultraviolet
limit for different values of $\mu _{R}-\mu _{L}$ leads always to the same
plateau structure. We observe a relatively sharp increase in $G$ for an 
\FIGURE{\epsfig{file=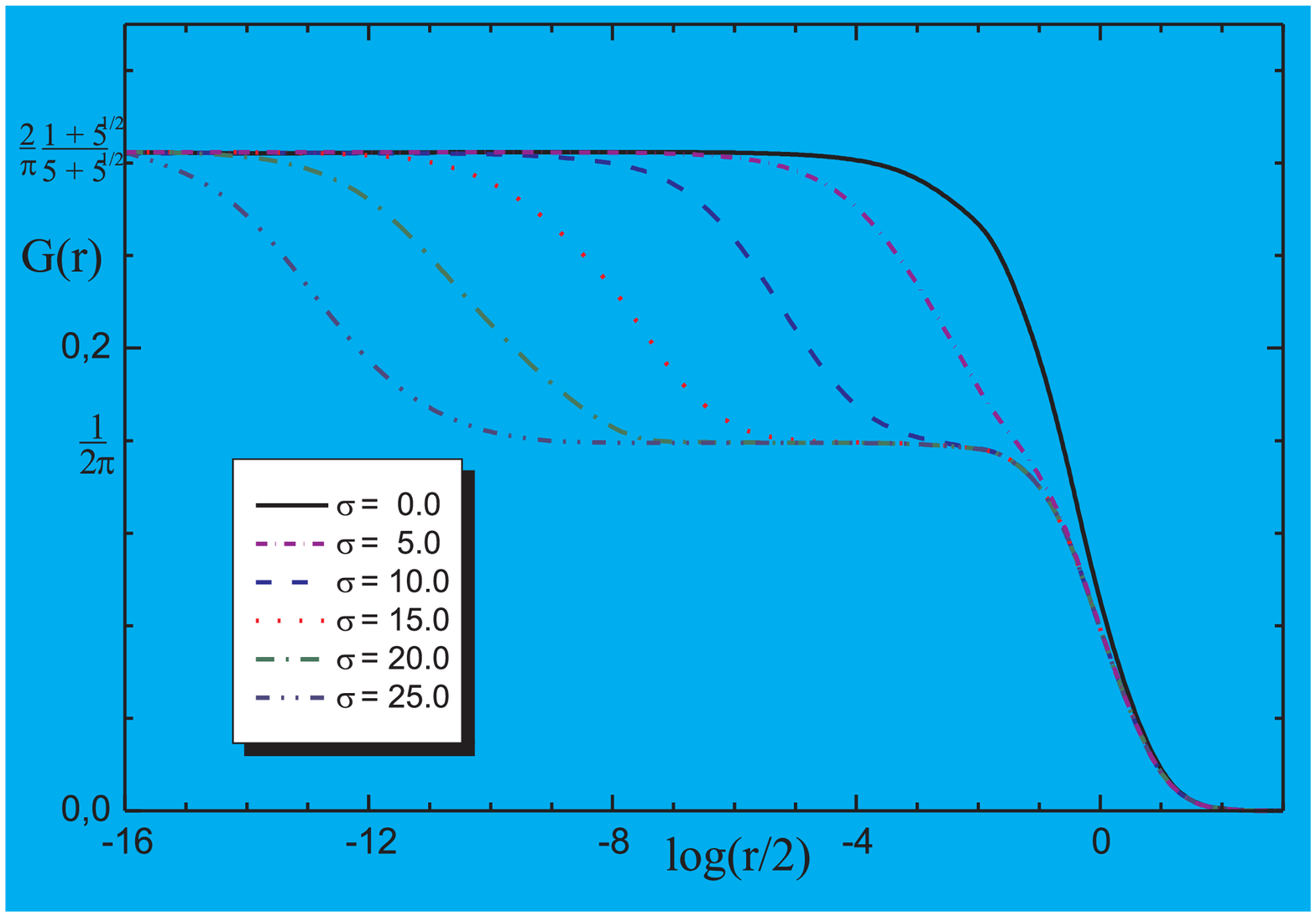,width=8.2cm,height=6.09cm}
\caption{Conductance for the $SU(3)_{2}$-HSG-model.}} \noindent energy scale 
$2\log r/2\sim -\sigma $ which corresponds to the onset of the unstable
particle. In other words, only when a certain energy scale necessary for the
excitation of the unstable particle is reached, the latter is formed and
participates in the conducting process. All this information is encoded in
the density $\rho _{i}^{r}(\theta ,r,\mu _{i})$. Computing now $\varepsilon
_{i}(\theta ,0,0)$ in a standard TBA fashion we predict the plateaux from (%
\ref{g0}) analytically at $1/2\pi $ and $2(1+\sqrt{5})/(5+\sqrt{5})\pi $.
The last plateau corresponds to the deep ultraviolet limit, whereas the
plateau at $1/2\pi $ coincides with the value (\ref{hh}) for a free Fermion
theory when taking $e/h=$ $1/2\pi .$ The reason is that the second plateau
in figure 5 develops in the region when $\ \sigma \gg -2\log r/2$, that is $%
\sigma $ very large. In that limit we have $\lim_{\sigma \rightarrow \infty
}S_{\pm \mp }(\theta )$ $=1$, such that the model becomes a free Fermion
theory.

\subsubsection{Resonances versus unstable particles}

\noindent In the light of the results of subsections 2.2.2 and 2.2.3 we can
draw the conclusion that resonances in a double defect system and the
presence of unstable particles may be described similarly \cite{CF7}.
Comparing figures 3(b) and 4, it is clear that the plateau structures
encountered do not differ much from each other. In particular, it seems that
the parameter $y$ in the double defect system and the resonance parameter $%
\sigma $ in the $SU(3)_{2}$-HSG model play similar roles. Let us investigate
more precisely these similarities, which from an intuitive point of view
appear rather natural.

In the context of theories possessing unstable particle in their spectra, a
very clear picture which explains the relatively sharp onset of the
conductance with increasing temperature can be provided. The temperature at
which this onset occurs, say $T_{C}$ can be related directly to the energy
scale at which the unstable particle is formed, since then it starts to
participate in the conducting process. The Breit-Wigner formula \cite{BW}
provides in this case the expressions for the mass $M_{\tilde{c}}$ and the
decay width $\Gamma _{\tilde{c}}$ of the unstable particle $\tilde{c}$.
Supposing that the particle $\tilde{c}$ is formed in the scattering process
between particles of types $i$ and $j$ of masses $m_{i},m_{j},$ this is
reflected by a pole in $S_{ij}(\theta )$ at $\theta _{R}=\sigma -i\bar{\sigma%
}$. Setting $\bar{\sigma}=\pi /2$, as corresponds to the model at hand, the
Breit-Wigner formula for large values of the resonance parameter $\sigma $
gives 
\begin{equation}
M_{\tilde{c}}\approx \frac{1}{\sqrt{2}}\sqrt{m_{i}m_{j}}\exp |\sigma
|/2\qquad \mathrm{and}\mathnormal{\qquad }\Gamma _{\tilde{c}}^{{}}\approx 
\sqrt{2m_{i}m_{j}}\exp |\sigma |/2\,.  \label{BW}
\end{equation}
Since a renormalization group flow is provided by mapping $M\rightarrow r\,M$%
, one observes that the quantity $M_{\tilde{c}}(r,\sigma )=rM=re^{\sigma /2}$
should remain invariant under the renormalization group flow. That means
that if $\ r_{1}$ is the onset energy for the unstable particle $\tilde{c}$
for $\sigma =\sigma _{1}$ and $r_{2}$ is the onset energy for $\sigma
=\sigma _{2}$, the conductance must satisfy the following scaling law 
\begin{equation}
G(r_{1},\sigma _{1})=G(r_{2},\sigma _{2})\,\,\,\quad \mathrm{for\quad }%
\,r_{1}e^{\sigma _{1}/2}=r_{2}e^{\sigma _{2}/2}.  \label{M12}
\end{equation}
This means we can control the position of the onset in the conductance by $%
M_{\tilde{c}}(r,\sigma )$.

Analyzing now the scaling behaviour of the conductance for the double defect
system studied in subsection 2.2.2 we find 
\begin{equation}
G(r_{1},y_{1})=G(r_{2},y_{2})\,\,\,\quad \mathrm{for}\quad \frac{r_{1}}{y_{1}%
}=\frac{\,r_{2}}{y_{2}},  \label{scal}
\end{equation}
Then the comparison with (\ref{M12}) suggests that we can formally relate
the distance between the two defects to the resonance parameter as $\sigma
=2\ln ($const$/y)$. However, despite the fact that the net result with
regard to the conductance is the same, the origin of the onset is different.
Whereas for the HSG-model it resulted from a change in the density
distribution function $\rho _{i}^{r}(\theta ,r,\mu _{i})$ it is now
triggered by the structure of $\,\left| T^{\mathbf{\alpha }}\left( \theta
\right) \right| .$ Since for the free Fermion the function $\rho _{i}^{r}$
keeps its overall shape and just translates as the temperature is changed,
the onset of the conductance occurs when the maxima of $\,\left| T^{\mathbf{%
\alpha }}\left( \theta \right) \right| $ are reached. By analyzing the
concrete expression of $\,T^{\mathbf{\alpha }}\left( \theta \right) $ for
the energy operator defect (see part I) it is easy to verify that for a
double defect 
\begin{equation}
T^{\alpha _{1}\alpha _{2}}(\theta =\ln \left[ \frac{(2n+1)\pi }{y}\right]
)\approx 1\,\quad \mathrm{for}\quad n\in \Bbb{Z.}  \label{peak}
\end{equation}
Drawing an analogy to the scattering matrix of the HSG-model the values of $%
\theta $ for which $T^{\alpha _{1}\alpha _{2}}(\theta )$ is maximal play the
same role as the value $\theta _{R}=\sigma -i\pi /2$ corresponding to the
resonance pole of the S-matrix. In that sense we can make the identification 
$\sigma _{n}=\ln \left[ (2n+1)\pi /y\right] .$ There are however some
differences between both systems, since in the case of the $SU(3)_{2}$-HSG
model the onset of the conductance is due to a single unstable particle,
whereas for the double defect system the same effect can be attributed to
several maxima of the transmission probability. The other important
difference is that $y$ is now a measurable quantity, so that the ``mass'' of
the resonances can be experimentally accessible.

\subsubsection{Multiple plateaux}

\noindent Up to now, we have observed that we always obtain essentially two
plateaux in the conductance, no matter how many ($\geq 2$) and what type of
defects we implement. The natural 
\FIGURE{\epsfig{file=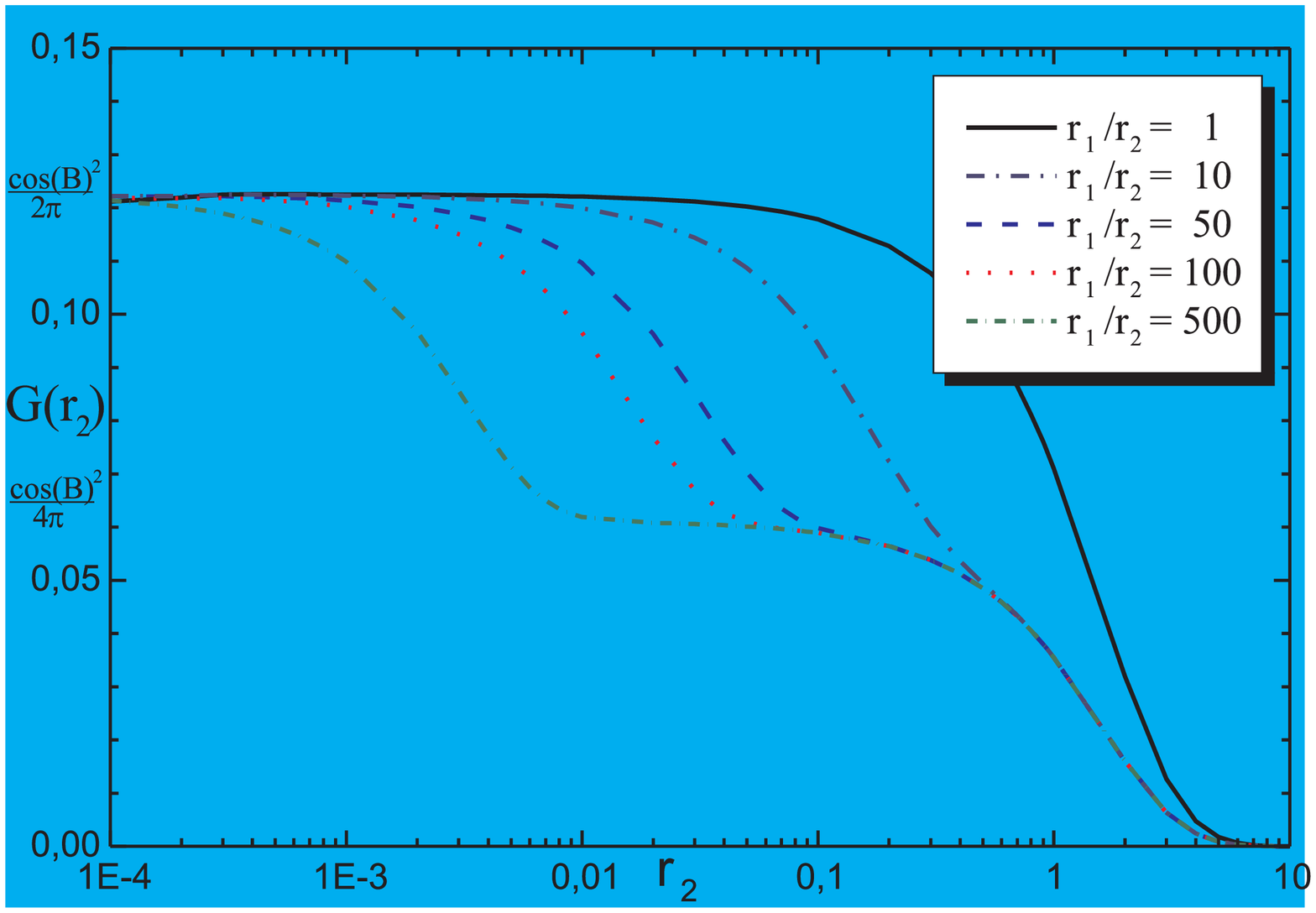,width=8.2cm,height=6.09cm}
\caption{Conductance $G(r_{2})$ for the complex
free Fermion with the energy operator defects as a function of the inverse
temperature $r_{2}$, for fixed effective coupling constant $B=0.5$
and varying temperature ratios in the two halves of the wire.}} \noindent
question arising at this point is whether it is possible to have a set up
which leads to a more involved plateaux structure. It is clear that if we
had many defects in a row separated far enough from each other such that the
relaxation time of the passing particles is so large that they could be
treated as single rather than multiple defects, then any desired type of
multiple plateau structure could be obtained. In this case the conductance
is simply the sum of the expressions one has for each defect independently.
Recalling the origin of the different plateaux, there is another slightly
less obvious option. The density distribution function $\rho ^{r}$ is a
peaked function of the rapidity and if the resonances in $T^{\mathbf{\alpha }%
}\left( \theta \right) $ would be separated far enough, such that they are
resolved by $\rho ^{r}$, we would also get a multiple plateaux pattern.
However, tuning the distance between the defects or the coupling constant
will merely translate the position of the resonances in the rapidity
variable or change their amplitudes, respectively (see section 2, part I).
Therefore the last option left is to change the $\rho ^{r}$s, which is
possible by varying the temperature. Choosing now a configuration as in
figure 2 with different temperatures $T_{1}$ and $T_{2} $, one can
``create'' a second plateau at half the height of the original one. The
reason for this is simply that the cooled half of the wire will cease to
contribute to the conductance as can be directly deduced from (\ref{gg}). We
depict the results of our computations in figure 5. From this it also
obvious that if we only cool the fraction $x$ of the wire, the lowest
plateau will be positioned at the height $x$ times the height of the upper
plateau. Thus, by combining these different configurations, i.e., different
temperatures or defects, we could produce any desired plateau structure.

\section{Conclusions and open problems}

\vspace{-0.2cm} \setcounter{equation}{0}

In this section I will present the main conclusions of my talk and also of
part I, since the main aim of this work was actually to compare the two
theoretical descriptions presented in the two parts. In our work we have
exploited the special features of 1+1 dimensional integrable quantum field
theories in order to compute the DC conductance in an impurity system. For
this purpose several non-perturbative techniques have been used. As the main
tools we employed the thermodynamic Bethe ansatz in a Landauer transport
theory computation and the form factor expansion in the Kubo formula.

\emph{The comparison between the Kubo formula (1.1; part I) and the Landauer
formula (\ref{I11}) yields in particular an identical plateau structure for
the DC conductance in the ultraviolet limit. }

We have explained to what extend integrability can be exploited in order to
determine the reflection and transmission amplitudes through a defect.
Unfortunately, for the most interesting situation in this context, namely
when $R/\tilde{R}$ and $T/\tilde{T}$ are simultaneously non-vanishing, the
Yang-Baxter bootstrap equations narrow down the possible bulk theories to
those which possess rapidity independent scattering matrices \cite{DMS,CFG}.
By means of a relativistic potential scattering theory we computed for
several types of defects the $R/\tilde{R}$s and $T/\tilde{T}$s, thus
enlarging the set of examples available at present. We confirm that for real
potentials parity is preserved, but otherwise essentially all possible
combinations of parity breaking can occur. From the knowledge of the single
defect amplitudes the multiple defect amplitudes, which exhibit the most
interesting physical behaviours, can be computed in a standard fashion \cite
{CT,Merz}.

We have newly proposed a Kubo formula \cite{kubo} which accommodates the
situation when defects are present (1.1; part I). We evaluated the
current-current correlation functions occurring in there by means of a
non-perturbative method based on integrability, namely the bootstrap form
factor approach \cite{KW,Smir}. We provide closed formulae which solve
explicitly the defect recursive equations involving any arbitrary number of
particles. We predict the plateaux in the conductance as a function of the
temperature analytically.

We newly formulated the TBA equations for a defect with simultaneously
non-vanishing reflection and transmission amplitudes. We indicate how these
equations can be used to compute various thermodynamic quantities, which
are, however, most interesting only when considered per unit length. By
means of the TBA we compute the density distribution functions and use them
to evaluate the Landauer conductance formula (\ref{I11}) for various defects
in a complex free Fermionic theory. Also in this case, we predict
analytically the most prominent features in the conductance as a function of
the temperature, i.e. the plateaux.

There exist various investigations, e.g., \cite{WA,FLS,FLS2,pureT} for
conformal (massless) theories with defects, which exploit the original
folding idea of Wong and Affleck \cite{WA}. The idea is that a conformal
field theory with a \emph{purely} transmitting or reflecting defect can be
mapped into a boundary theory, i.e. a theory living in half space, which has
the advantage that the full  restriction of modular invariance can be
exploited in the construction of boundary states as pioneered by Cardy \cite
{Cardy}. Apparently the folding procedure could lead to non-trivial
solutions for the reflection and transmission amplitudes starting with a
purely reflecting or transmitting theory. However, one should stress that
the folding is carried out on the basis of the field content of the
conformal field theory, whereas our analysis is based on a particle
description, namely we take the ZF-algebra as our starting point. Therefore,
the transmission and reflection amplitudes obtained by means of the folding
technique can not be compared with the objects we study here, even in the
conformal limit.

In this context, there are several interesting open issues. Most challenging
is to treat in full generality the massive and temperature dependent case of
(1.1; part I). Unfortunately, the formulation of non-perturbative methods
does not yet cover that situation \cite{CF8} and it remains to be clarified
how the form factor bootstrap program for the computation of two-point
functions can be extended to that case. It would be further interesting to
compute thermodynamic quantities per unit length by means of the TBA and to
develop methods for the systematic classification of integrable defects.

Proceeding further in our investigation of the applications of integrable
models to the description of realistic physical systems, we have established
that \emph{coupling an impurity in a quantum wire to an external
monochromatic electromagnetic field leads to high harmonic generation \cite
{CFF}}. Harmonic generation i.e., the emission of multiples of the incoming
frequency when a system is coupled to a monochromatic field, has been widely
studied in the context of atomic physics. However, up to now there were no
results for solid state materials. The concrete system we have studied is a
quantum wire described by means of the Dirac equation doped with a defect
which couples minimally to an external field of frequency $\omega $.
Considering separately the situations corresponding to a single and a double
defect system we observed that, for the particular type of defect treated,
only even multiples of the incoming frequency are emitted for the single
defect, whereas all even and odd multiples are generated for the double
defect system. These features are observed both in the Fourier expansion of
the transmission probability through the defect and in the emission spectrum
of the dipole momentum. It would be extremely interesting to confirm our
findings experimentally.

\acknowledgments

We would like to thank the organizers of this workshop for the opportunity
to present these talks, for financial support and for making the celebration
of the $50^{\mathrm{th}}$ anniversary of the Instituto de F\'{\i}sica 
Te\'orica (IFT) a very enjoyable and
rewarding event. In addition, we would like to thank Carla Figueira de
Morisson Faria (Max Born Institut Berlin) and Frank G\"{o}hmann (Universit%
\"{a}t Bayreuth) for their participation in \cite{CFF} and \cite{CFG},
respectively. We are also grateful to the Deutsche Forschungsgemeinschaft
(Sfb288) for financial support.

\end{document}